\documentclass{WileyMSP-template}
\usepackage{graphicx}
\usepackage{dcolumn}
\usepackage{multicol}
\usepackage{braket}
\usepackage{float}
\usepackage{bm}
\usepackage{amsmath}
\usepackage{cite}
\usepackage{hyperref}
\usepackage{xcolor}
\usepackage{amssymb}
\usepackage{gensymb}
\usepackage{dirtytalk}
\usepackage [english]{babel}
\usepackage [autostyle, english = american]{csquotes}
\usepackage{textcomp}

\usepackage{cuted}
\usepackage{caption}
\usepackage{tabularx}
\usepackage{booktabs}
\usepackage{multirow}

\MakeOuterQuote{"}

\begin{document}

\rhead{\includegraphics[width=2.5cm]}

\title{Geometric phase-assisted simple phase compensation enabling quantum key distribution using phase-shifted Bell states
}

\maketitle

\vspace{12pt}
\author{Ayan Kumar Nai},
\author{G. K. Samanta}





\vspace{12pt}
\begin{affiliations}
Ayan Kumar Nai\\
Photonic Sciences Lab., Physical Research Laboratory, Ahmedabad 380009, Gujarat, India\\
Indian Institute of Technology Gandhinagar, Palaj, Gandhinagar, 382055, Gujarat, India\\
Email Address: ayankrnai@gmail.com\\
\vspace{12pt}

G. K. Samanta\\
Photonic Sciences Lab., Physical Research Laboratory, Ahmedabad 380009, Gujarat, India\\
\vspace{12pt}

\end{affiliations}



\justifying
\begin{abstract}

Entanglement-based quantum key distribution (QKD) relies on the distribution of high-fidelity maximally entangled Bell states, typically generated via spontaneous parametric down-conversion (SPDC). In practical systems, unwanted relative phases arise from birefringence, pump-beam contributions, imperfect photon-pair generation, transmission through physical channels, and collection, transforming Bell states into phase-shifted states. This degrades interference visibility, increases the quantum bit error rate (QBER), and limits secure key generation. Conventional compensation techniques, such as birefringent crystals, interferometric stabilization, and spatial light modulators, are often impractical in real-world deployments. 
Here, we demonstrate a simple and versatile phase-compensation scheme that can be implemented at either the source or the receiver to eliminate arbitrary relative phases in Bell states. We theoretically and experimentally quantify the dependence of QBER in the BBM92 protocol on the relative phase and show that geometric-phase-based control can effectively restore entanglement quality. In a proof-of-concept experiment using a nondegenerate polarization Bell state, we achieve a fidelity exceeding 95$\%$ and reduce QBER below the 11$\%$ security threshold required for secure QKD. This robust approach enables practical phase control in entangled-photon systems and can be extended to time-bin QKD via time–polarization mapping, offering a promising route toward stable, low-QBER quantum communication.

\end{abstract}

\begin{multicols}{2}

\justifying
\section{Introduction}

Entanglement-based quantum key distribution (QKD) protocols rely critically on the stable generation and preservation of maximally entangled Bell states shared between distant parties \cite{ekert1991, bennett1992, gisin2010}. A major challenge in such systems is the introduction of unwanted phase factors that degrade entanglement visibility and increase the quantum bit error rate (QBER), thereby limiting secure key generation. Phase drift in entangled photon pairs generated through spontaneous parametric down-conversion (SPDC) arises from several sources, including birefringence of optical components, phase contributions from the pump beam \cite{Kwiat:99,kumar2025}, imperfections in nonlinear crystal generation \cite{kim2006phase, kim2019pulsed}, and transmission through practical environments \cite{wang2011degree}. Therefore, effective phase compensation is essential for maintaining continuous low-QBER operation in realistic QKD systems.

To date, various phase compensation techniques have been developed and implemented either at the pre-generation or post-generation stage \cite{feng2024pre}. Common approaches include the use of birefringent crystals in the pump beam \cite{zhou2013ultra, cialdi2011nonlocal}, phase-compensating crystals after generation \cite{zhou2013ultra, steinlechner2013}, dual-crystal configurations \cite{kwiat1995new, altepeter2005phase}, and specially engineered SPDC sources \cite{feng2023ent}. Additional methods such as interferometric techniques \cite{mo2005faraday}, spatial light modulation \cite{zhou2010phase, cialdi2011nonlocal}, and polarization control using fiber paddles \cite{patel2016quantum} have also been reported. However, many of these approaches provide static compensation and are insufficient for dynamically varying phases in practical scenarios. Active phase compensation is therefore essential for real-time monitoring and correction of phase drift \cite{tagliavacche2025, chen2008active}. Interferometric methods play a key role in estimating dynamic phase variations arising from relative path differences \cite{Lai:91, Joenathan:94, wyant2003}, although such phases are inherently noisy and time-dependent. In contrast, the geometric phase \cite{pancharatnam1956}, arising from the evolution of the polarization state on the Poincaré sphere, is intrinsically stable and path-independent, making it a powerful tool for phase control and compensation. Recent studies have shown that the geometric phase of a classical pump beam can be transferred to polarization-entangled photon pairs generated via SPDC, enabling tunable control over entanglement properties \cite{kumar2025}. By rotating a half-wave plate sandwiched between two quarter-wave plates (QHQ configuration), a controllable geometric phase is imprinted onto the biphoton state, allowing continuous transformation between Bell states and intermediate maximally entangled states. This phase, also known as the Pancharatnam–Berry phase \cite{pancharatnam1956, berry1987}, has found wide applications in classical and quantum optics, including frequency shifting \cite{simon1988}, quantum gates \cite{sjoqvist2008}, quantum metrology \cite{zhou2025}, and beam shaping \cite{conrads2025}, and can be readily implemented through polarization manipulation using wave plates \cite{simon1989}. Despite significant progress in QKD implementations beyond laboratory settings, relatively few studies have explored the effect of the relative phase of Bell states on the performance and security of entanglement-based QKD \cite{gisin2002quantum, scarani2009}, largely due to the lack of precise and stable phase control. Recently, we reported the generation of phase-shifted Bell states with continuously tunable relative phase using geometric phase \cite{kumar2025, Davis:25}.

Here, we present the theoretical and experimental study of QKD using phase-shifted Bell states. By introducing a geometric phase in the pump beam of an SPDC source implemented in a polarization Sagnac interferometer, we generate Bell states with controllable relative phase. Using the BBM92 protocol between two users, we measure the QBER as a function of the relative phase and identify the range of relative phases compatible with secure QKD. We further demonstrate that the induced phase can be effectively compensated at the detection stage using a geometric phase element, restoring low-QBER operation. Additionally, we show that unwanted phase acquired during generation or transmission can be corrected by applying a geometric phase at the pump and/or at the user end. These results establish geometric phase as a practical single-parameter tool for both engineering and correcting entangled states, significantly relaxing source optimization requirements and enabling robust, high-quality entanglement for realistic QKD implementations. 

\section{Theory}
The maximally entangled state generated via non-collinear, degenerate, type-0 phase matching in a periodically poled nonlinear crystal placed inside a polarization Sagnac interferometer \cite{Jabir:2017}, or via type-I phase matching using a dual thin-crystal SPDC configuration \cite{Kwiat:99}, can be written as

\begin{equation}
|\psi_{\phi}\rangle =  \frac{1}{\sqrt{2}}\left(|H\rangle_A |H\rangle_B \pm e^{i\phi}|V\rangle_A |V\rangle_B\right),
\label{Eq1}
\end{equation}
where the relative phase $\phi$ depends on several factors, including birefringence, dispersion, pump-induced phase fluctuations, imperfect photon-pair generation and collection, as well as environmental disturbances during transmission. The Eq. (\ref{Eq1}) reduces to the standard polarization-entangled Bell states $|\Phi^{\pm}\rangle$ when $e^{i\phi} = \pm 1$. For other values of $\phi$, the relative phase modifies the interference visibility in specific measurement bases, often leading to a reduced Bell parameter when measurements are performed in the conventional Bell basis \cite{Radhika:09}. However, the presence of a relative phase does not reduce the degree of entanglement. The state remains maximally entangled under local unitary transformations \cite{kumar2025}. Such states are therefore referred to as \textit{phase-shifted} Bell states \cite{Davis:25}.

Using simple mathematical steps \cite{kumar2025}, one can find the CHSH Bell's parameter \cite{clauser1969} of the phase-shifted Bell states as 

\begin{equation}
S = \left| \sqrt{2} \right| + \left| \sqrt{2} \cos \phi \right|
\label{Eq2}
\end{equation}

Again, using the relation between the entanglement visibility ($V$) and the CHSH Bell parameter ($S$) \cite{weihs1998violation}, we can express the entanglement visibility as a function of the relative phase $\phi$ as

\begin{align}
V &= \frac{S}{\left| 2\sqrt{2} \right|} \notag \\
&= \frac{\left| \sqrt{2} \right| + \left| \sqrt{2} \cos \phi \right|}{\left| 2\sqrt{2} \right|}\notag\\
&= \frac{1 + \left|  \cos \phi \right|}{2}
\label{Eq3}
\end{align}
Subsequently, the quantum bit error rate (QBER) in entanglement-based QKD \cite{scheidl2009feasibility, neumann2021model} using the phase-shifted Bell state can be written as,

\begin{equation}
QBER = \frac{1 - V}{2} = \frac{1 - \left|  \cos \phi \right|}{4}
\label{Eq4}
\end{equation}
It is evident from Eq. (\ref{Eq4}) that the QBER attains its minimum value of $0\%$ when $\phi = 0^\circ$. Therefore, entanglement-based QKD is conventionally implemented using standard Bell states to provide a larger tolerance against other experimental imperfections that contribute to the QBER, thereby ensuring secure key distribution within the acceptable QBER threshold. On the other hand, the QKD using phase-shifted Bell state can have a maximum QBER due to the source as 25\% for $\phi = 90\degree$. A relative phase of $\phi$ = 0.977 rad, the QBER is 11$\%$. Even at the source, the generated state is a Bell state without relative phase; however, due to the presence of different optical components in the experiment, it transforms into a phase-shifted Bell state with a higher QBER. To overcome such a phase shift in the Bell state in a practical demonstration of entanglement-based QKD, we can apply the same amount of phase in opposite sign to the generated state in two ways: at the pump beam and direct transfer to the generated state, or at the detection time by one of the communicating parties in a controlled way. Since the geometric phase is stable and robust against any external perturbation, we can use it as a simple controlled parameter for the automatic transfer of the phase-shifted Bell state to the standard Bell state. 

The transfer matrix of the QHQ setup, where the quarter-wave plate (QWP, Q), half-wave plate (HWP, H), and quarter-wave plate (QWP, Q), sequentially positioned with their fast axes oriented at angles 45$\degree$, $\theta ^P_H$, and 45$\degree$, with respect to the vertical, respectively, can be simplified to 

\begin{equation}
\begin{aligned}
T = QHQ &= 
\begin{pmatrix}
e^{2i\theta ^P_H} & 0 \\
0 & -e^{-2i\theta ^P_H}
\end{pmatrix}
\\[1ex] 
\end{aligned}
\label{Eq5}
\end{equation}
Here, the geometric phase, $\phi^P_g$ (= 2$\theta ^P_H$), governed by a single parameter (the angle of the half-wave plate), can introduce any arbitrary amount of phase. 

As presented in Ref. \cite{kumar2025}, the transfer of the geometrical phase (GP) of the classical pump beam of diagonal polarization on propagation through the GP-setup with HWP at an angle, $\theta ^P_H$ transforms into the polarization state of $\frac{1}{\sqrt{2}}(e^{2i\theta_H^P}|H\rangle - e^{-2i\theta_H^P)}|V\rangle )$ and generate the entangled state as

\begin{equation}
|\psi_{\phi}^{f}\rangle \equiv \frac{1}{\sqrt{2}}\left(|H\rangle_A |H\rangle_B - e^{i(\phi_{un} + 2\phi^P_g)}|V\rangle_A |V\rangle_B\right)
\label{Eq6}
\end{equation}
here, $\phi_{un}$ is the unknown phase acquired by the entangled state during the generation or distribution process. We have ignored the global phase.

As evident from Eq. \ref{Eq6}, the entangled state shared between the parties is a phase-shifted Bell state with an unknown phase $\phi_{un}$ and a controllable geometric phase $\phi^P_g$. Now, if either Alice or Bob employs a QHQ setup with the HWP oriented at an angle $\theta^{AB}_H$ at their detection station before the beam splitter, then using Eq. \ref{Eq5} and Eq. \ref{Eq6}, the shared entangled state can be written as

\begin{align}
|\psi_{\phi}^{f}\rangle &= \frac{1}{\sqrt{2}} \Big(
|H\rangle_A e^{i\phi^{AB}_g}|H\rangle_B - \notag \\
&\quad e^{i(\phi_{un} + 2\phi^P_g)} 
|V\rangle_A e^{-i\phi^{AB}_g}(-1)|V\rangle_B \Big) \notag \\
&\equiv \frac{1}{\sqrt{2}} \Big(
|H\rangle_A |H\rangle_B \notag + \notag \\
&\quad e^{i(\phi_{un} + 2\phi^P_g - 2\phi^{AB}_g)}
|V\rangle_A |V\rangle_B \Big)
\label{Eq7}
\end{align}

Here, the global phase has been ignored in both Eq. \ref{Eq6} and Eq. \ref{Eq7}. The final state shared between two parties given by Eq. \ref{Eq7} becomes a maximally entangled Bell state,
without any relative phase, when the following condition is satisfied,

\begin{gather}
e^{i(\phi_{un} + 2\phi^P_g - 2\phi^{AB}_g)} = 1 \notag \\
\phi_{un} + 2\phi^P_g - 2\phi^{AB}_g = 2n\pi \notag \\
\phi_{un} = -2\phi^P_g + 2\phi^{AB}_g
\label{Eq8}
\end{gather}

\noindent where $n = 0, 1, 2, \dots$ is an integer. It follows from Eq. \ref{Eq8} that the arbitrary relative phase of the Bell state, acquired during generation or distribution, can be compensated by introducing an appropriate geometric phase in the pump beam (for $\phi^{AB}_g = 0$), at the detection stage by the communicating parties (for $\phi^P_g = 0$), or by a combination of both in QKD.

\section{Experiment}
\begin{figure*}[ht]
    \centering
    \includegraphics[width=\linewidth]{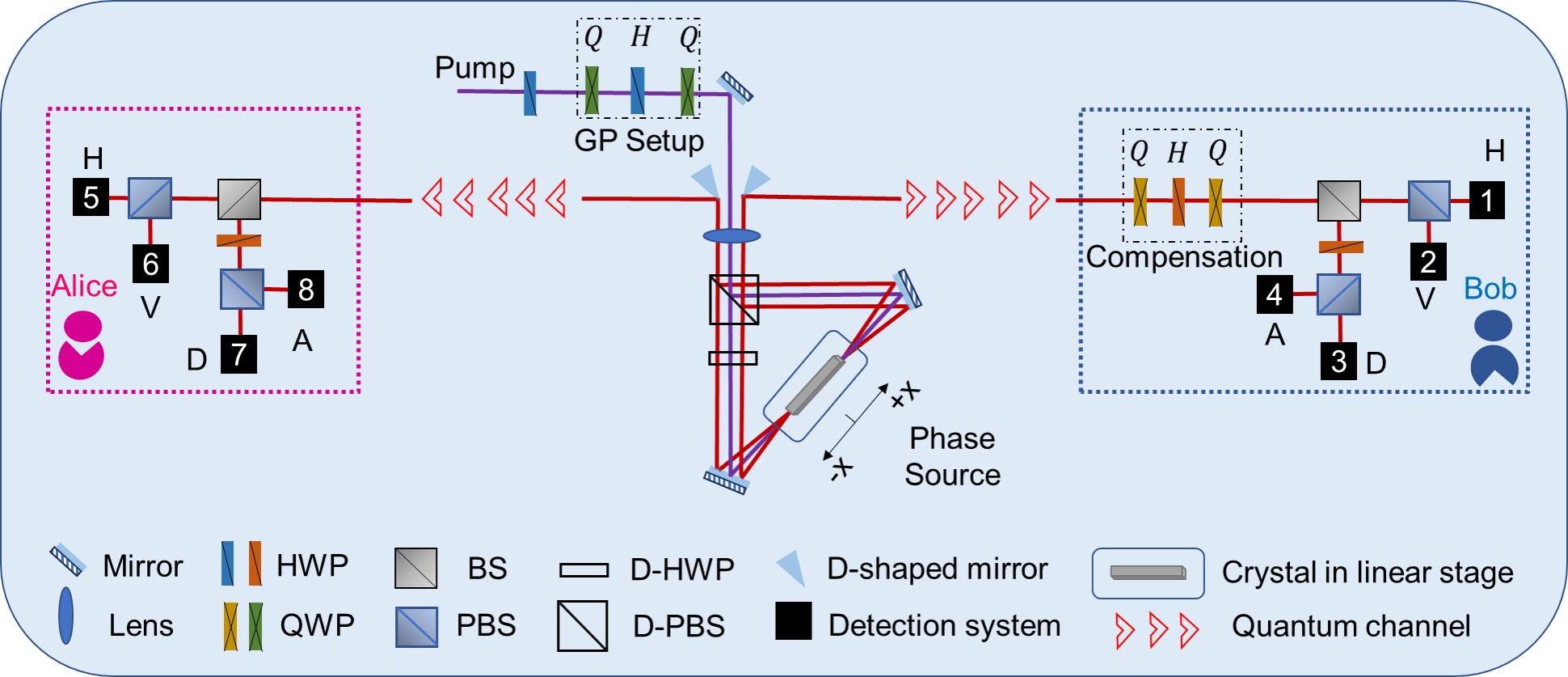}
    \caption{The schematic representation of the experimental setup of the BBM92 key distribution with active phase compensation. Pump: 405 nm, cw diode laser, GP: geometric phase, HWP (H): half-wave plate, QWP (Q): quarter-wave plate, BS: beam-splitter, PBS: polarising beam-splitter, Lens: plano-convex lens of focal length 150 mm, D-HWP: Dual half-wave plate, D-PBS: dual polarising beam-splitter, Crystal: 10 mm long PPKTP crystal of single grating period in temperature oven kept on linear stage,   1-8: single photon counting modules.}
    \label{Figure1}
\end{figure*}

The schematic of the experimental setup for verifying effective phase compensation of the phase-shifted Bell state in QKD is shown in Fig. \ref{Figure1}. Entangled photons are generated using a polarization-based Sagnac interferometer geometry \cite{Jabir:2017, Singh:2023}, consisting of a dual-wavelength (405 nm and 810 nm) polarizing beam splitter (D-PBS) cube, a half-wave plate (D-HWP), and two high-reflecting plane mirrors. A 10 mm-long periodically poled potassium titanyl phosphate (PPKTP) crystal, with a grating period $\Lambda = 3.245~\mu$m and a clear aperture of $1\times2~\text{mm}^2$, is placed at the center between the mirrors, equidistant from the D-PBS along both clockwise (CW) and counterclockwise (CCW) directions. The crystal produces noncollinear, degenerate, type-0 ($e \to e + e$) phase-matched spontaneous parametric down-conversion (SPDC) at 810 nm. It is housed in a temperature-controlled oven stabilized to within $\pm0.1^\circ$C, with a maximum temperature of $200^\circ$C.
The pump source is a continuous-wave, single-frequency laser diode operating at 405.03 nm with a linewidth of approximately 20 MHz, focused at the crystal center using a plano-convex lens of focal length $f = 150$ mm. A geometric phase (GP) is imprinted on the pump beam using a sequence of a quarter-wave plate (QWP), a half-wave plate (HWP), and another QWP, with fast axes oriented at $45^\circ$, $\theta_H^P$, and $45^\circ$, respectively. The input pump beam is first converted to diagonal polarization using an HWP. Upon passing through the GP setup, it acquires a controllable relative phase determined by the HWP angle $\theta_H^P$, following the transfer matrix given in Eq. \ref{Eq5}. This phase-controlled diagonal pump generates entangled photons with an annular spatial distribution \cite{Jabir:2017, Singh:2023}. The SPDC ring is divided into diagonally opposite sections using two D-shaped mirrors. The generated entangled state shared between two parties is given by \cite{kumar2025},
$\ket{\Phi^{-}} \equiv \frac{1}{\sqrt{2}} \left(\ket{H_AH_B} - e^{i(\phi_{un} + 2\phi_g^P)}\ket{V_AV_B}\right)$,
originating from the pump polarization state $\frac{1}{\sqrt{2}}\left(e^{i\phi_g^P}|H\rangle_p - e^{-i\phi_g^P}|V\rangle_p\right)$ as in Eq. \ref{Eq6}. Here, $\phi_{un}$ arises due to the asymmetric position of the crystal with respect to the dual HWP, and $\phi_g^P = 2\theta_H^P$ is the geometric phase introduced in the pump.

To implement BBM92-based QKD, Alice and Bob use standard polarization analysis setups. Each setup consists of a 50:50 non-polarizing beam splitter (BS) cube, a polarizing beam splitter (PBS) cube for Z-basis (H/V) projection in the transmitted arm, and a combination of an HWP (oriented at $22.5^\circ$) and PBS for X-basis (D/A) projection. The projected photons in H, V, D, and A polarization states are detected using four fiber-coupled single-photon counting modules (SPCMs), numbered 1–4 for Bob and 5–8 for Alice. These detectors are connected to a time-to-digital converter for time stamping, recording, and further analysis. For phase compensation of the entangled state, Bob (for example) employs an additional GP setup before the BS, identical to that used for the pump. By adjusting the HWP angle $\theta_H^{AB}$, a geometric phase $\phi_g^{AB} = 2\theta_H^{AB}$ is introduced, resulting in the final compensated quantum state as described in Eq. \ref{Eq7}.

\section{Results and discussions}
\label{RnD}

First, we studied the effect of a phase-shifted Bell state on the QBER in QKD between Alice and Bob. In our previous report \cite{kumar2025}, we observed that the geometric phase of the pump can be directly transferred to the SPDC photons, enabling control of the relative phase of the Bell state. We use the same principle here to generate a phase-shifted Bell state using a GP setup on the pump beam. If we assuming the phase acquired during generation and distribution process of the entangled state is $\phi = 0$, then according to Eq. \ref{Eq6}, the geometric phase induced phase-shifted Bell state can be written as $\ket{\Phi^{-}} = \frac{1}{\sqrt{2}} \left(\ket{H_AH_B} - e^{i 4\theta_g^P}\ket{V_AV_B}\right)$, where $\phi_g^P = 2\theta_g^P$ is the geometric phase controlled by the HWP angle $\theta_g^P$. The crystal is kept equidistant from the D-PBS to avoid additional relative phase contributions, so the phase shift arises solely from $\theta_g^P$.

\begin{figure}[H]
    \centering
    \includegraphics[width=\linewidth]{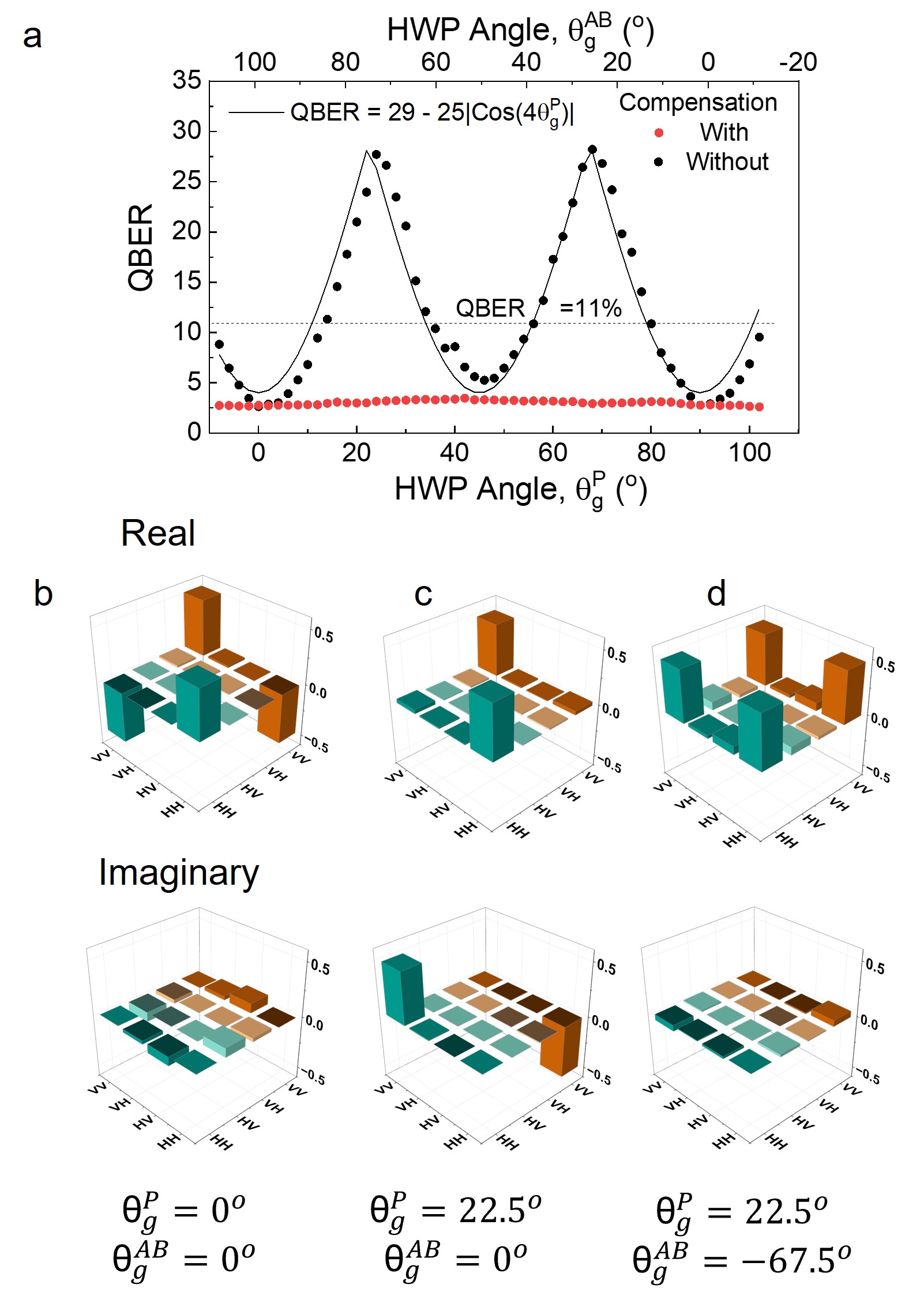}
    \caption{(a) Variation of QBER as a function of HWP angle inside the pump QHQ. Real and imaginary parts of experimentally reconstructed density matrix (b) at HWP angle $\theta_H^p = 0\degree$ (C) at HWP angle $\theta_H^p = 22.5\degree$ (d) at combined HWP angles $\theta_H^p = 22.5\degree$ \& $\theta_H^B = -67.5\degree$.}
    \label{Figure2}
\end{figure}

Keeping the projection systems of Alice and Bob unchanged and using the BBM92 protocol, we measured the QBER as a function of $\theta_g^P$, with the results shown in Fig. \ref{Figure2}. As evident from Fig. \ref{Figure2}(a), the variation of  $\theta_g^P$ from $0^\circ$ to $100^\circ$ at a pump power of 3 mW, results in the QBER oscillation from a minimum of $2.63\%$ at $0^\circ$ and $90^\circ$ to peaks of $\sim 28\%$ at $24^\circ$ and $66^\circ$, with an intermediate minimum of $5.2\%$ at $46^\circ$. The oscillation repeats beyond $90^\circ$. To understand this phase dependence, we fit (black line) Eq. \ref{Eq4} to the experimental data (black circle)  in the functional form of $\text{QBER} = a - b \left| \cos(4\theta_H^P) \right|$, and obtained $a = 29$ and $b = 25$, close to the theoretical values $a = b = 25$. The small shift in the angle of HWP in the experimental and theoretical results can be attributed to the resolution of the rotation mount used in the experiment. Putting the value of $\theta_H^P$ in the phase-shifted Bell state we can find that the minimum QBER is obtained for the Bell states $\frac{1}{\sqrt{2}} (\ket{H_AH_B} \mp \ket{V_AV_B})$, while the maximum QBER ($\sim 28\%$) occurs at $\theta_H^P = 22.5^\circ$ and $67.5^\circ$, corresponding to $\frac{1}{\sqrt{2}} (\ket{H_AH_B} \mp i\ket{V_AV_B})$, which exceed the $11\%$ security threshold and are therefore unsuitable for QKD. Considering the upper bound of the QBER for secure QKD, we can use phase-shifted Bell state with relative phase $\pm 0.31\pi$. However, in such a case, there will be no room for any external perturbation influencing QBER. 

Now, if a GP setup is used at the detection stage (Alice or Bob) to compensate the relative phase (Fig. \ref{Figure1}), the state (Eq. \ref{Eq7}) becomes $\ket{\Phi^{-}} = \frac{1}{\sqrt{2}} \left(\ket{H_AH_B} - e^{-i (4\theta_g^P - 4\theta_g^{AB})}\ket{V_AV_B}\right)$, where $\theta_g^{AB}$ is the HWP angle of the GP setup at the receiver. This reduces to a Bell state when $\theta_g^P = \theta_g^{AB}$ or $-(90^\circ - \theta_g^{AB})$, corresponding to parallel or orthogonal fast axes. Experimentally, after introducing a compensating GP setup at the user, the QBER remains nearly constant at $\sim 3\%$ (red dots) for all $\theta_g^P$ (Fig. \ref{Figure2}), confirming phase correction. The angles satisfy $\theta_g^{AB} = \theta_g^P$ or $-(90^\circ - \theta_g^P)$; results are shown for the latter case. We also reconstructed the states using quantum state tomography for ($\theta_g^P$, $\theta_g^{AB}$) = $(0^\circ, 0^\circ)$, $(22.5^\circ, 0^\circ)$, and $(22.5^\circ, -67.5^\circ)$ (Fig.~\ref{Figure2}(b--d)). The QBER is minimum for $\frac{1}{\sqrt{2}} (\ket{H_AH_B} - \ket{V_AV_B})$, maximum for $\frac{1}{\sqrt{2}} (\ket{H_AH_B} - i\ket{V_AV_B})$, and, after compensation, the state transforms to $\frac{1}{\sqrt{2}} (\ket{H_AH_B} + \ket{V_AV_B})$, with fidelity $\sim 98\%$.
From this study, it is evident that the geometric phase can serve as a control parameter to both transform a Bell state into a phase-shifted Bell state and compensate for unwanted phase acquired during practical implementations.

\begin{figure}[H]
    \centering
    \includegraphics[width=\linewidth]{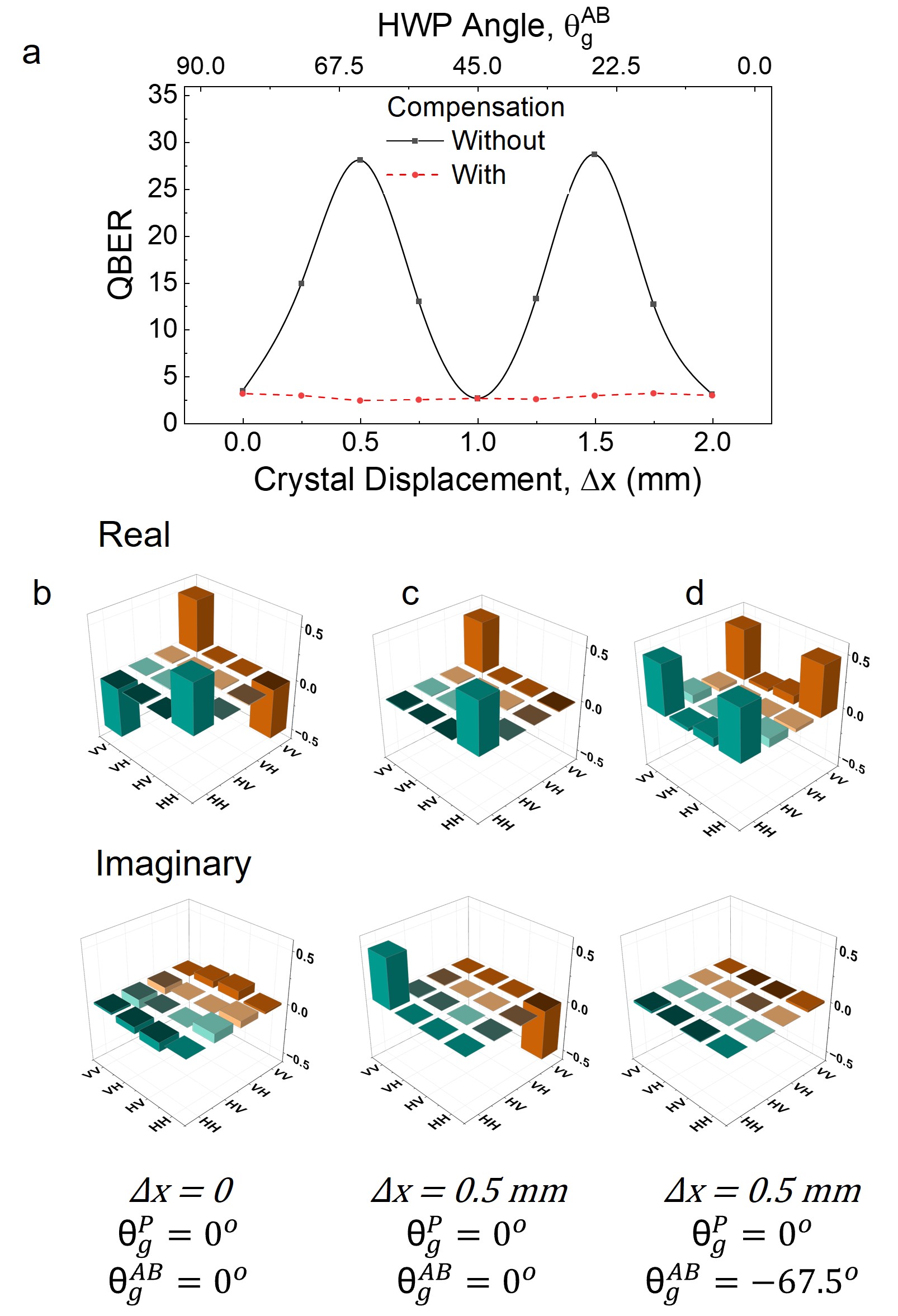}
    \caption{(a) Variation of QBER as a function of crystal position inside the Sagnac loop. Real and imaginary parts of experimentally reconstructed density matrix (b) at crystal position = 0 mm (C) at crystal position = 0.5 mm, and (d) at crystal position = 0.5 mm and $\theta_H^{AB} = -67.5\degree$.}
    \label{Figure3}
\end{figure}

After fully characterizing the phase-shifted Bell state and phase compensation using the geometric phase, we applied the same technique to measure and compensate the unknown phase in the QKD system. Keeping the HWP angles fixed at $\theta_H^P = \theta_H^{AB} = 0^\circ$, we measured the QBER while translating the crystal along the beam path inside the Sagnac interferometer. Since the crystal is initially positioned symmetrically with respect to the D-PBS, it can be displaced by a distance $\Delta x$ on either side of this point. As the phase is unknown, the HWPs were mounted on motorized rotation stages with a minimum step of 25 arcsec. The crystal was translated in increments of 0.25 mm, and the corresponding QBER was measured (Fig. \ref{Figure3}). As shown in Fig. \ref{Figure3}(a), the QBER oscillates between a minimum of $\sim 3\%$ and a maximum of $\sim 28\%$ as $\Delta x$ increases from 0 to 2 mm, similar to the behavior observed in Fig. \ref{Figure2}(a). Further displacement leads to a repetition of the oscillation, although the overall counts decrease due to reduced photon generation from lower parametric gain under asymmetric focusing conditions in the crystal.

We reconstructed the states using quantum state tomography for ($\theta_g^P$, $\theta_g^{AB}$) = $(0^\circ, 0^\circ)$ at different values of $\Delta x$. The density matrices shown in Fig. \ref{Figure3}(b,c) indicate that the state evolves from a Bell state to a phase-shifted Bell state with crystal displacement. The minimum QBER corresponds to $\frac{1}{\sqrt{2}} (\ket{H_AH_B} \mp \ket{V_AV_B})$ at $\Delta x = 0$, 1 mm, and 2 mm, while the maximum QBER corresponds to $\frac{1}{\sqrt{2}} (\ket{H_AH_B} \mp i\ket{V_AV_B})$ at $\Delta x = 0.5$ mm and 1.5 mm. The fidelity of the generated states exceeds 98$\%$ for all $\Delta x$. Further, we compensated the relative phase to maintain a minimum QBER for any crystal displacement by adjusting the HWP angles. For this study, $\theta_g^P$ was fixed at $0^\circ$, and $\theta_g^{AB}$ was varied to minimize the QBER. As shown by the red dots in Fig. \ref{Figure3}(a), the QBER remains nearly constant at $\sim 3\%$ for all $\Delta x$. Similar compensation is achieved by varying $\theta_g^P$ instead. Finally, we performed quantum state tomography for the state corresponding to maximum QBER at $\Delta x = 0.5$ mm after phase compensation using ($\theta_g^P$, $\theta_g^{AB}$) = $(0^\circ, -67.5^\circ)$. As shown in Fig. \ref{Figure3}(d), the reconstructed state is $\frac{1}{\sqrt{2}} (\ket{H_AH_B} + \ket{V_AV_B})$ with a fidelity of $\sim 98\%$. This study also highlights an important aspect of Bell-state generation, which is often affected by multiple experimental parameters. Since the relative phase can be easily compensated using the geometric phase at the pump or by the user through a single control parameter, the challenge of generating a perfect Bell state by simultaneously controlling all source parameters can be effectively avoided.

We summarize the results in Table \ref{tab:final}. As evident, the initial phase-shifted Bell state (second column), generated either through controlled geometric phase at the pump or from an unknown phase (first column) acquired during generation and distribution in the QKD implementation, exhibits a state fidelity of $>95\%$ (third column). By suitable optimization of the geometric phase implemented by the user (fifth column), the state is transformed into a Bell state (sixth column) with a fidelity exceeding $95\%$ (seventh column), in accordance with Eq. \ref{Eq8}. We have also tabulated the QBER values for the QKD using Bell states with different relative phases (fourth column) and compensated phase (eighth column). The results are in close agreement with the theoretical values.

\begin{table*}[t]
\centering
\caption{Summery of the transformation of phase-shifted Bell state using a single parameter for }
\label{tab:final}
\renewcommand{\arraystretch}{1.2}
\setlength{\tabcolsep}{6pt}

\begin{tabular}{c c c c| c c c c}
\toprule
\multicolumn{4}{c}{\textbf{Without Compensation}} 
& \multicolumn{4}{c}{\textbf{With Compensation}} \\
\cmidrule(lr){1-4} \cmidrule(lr){5-8}

$\theta_H^P$/$\Delta x$ & Initial States & Fidelity & QBER
& $\theta_H^{AB}$ & Compensated States & Fidelity & QBER\\
\midrule

$0^\circ$/0   & $\frac{1}{\sqrt{2}}(HH - VV)$   & 98\% / 98\%
& 2.73/3.4\% & $-90^\circ$   & $\frac{1}{\sqrt{2}}(HH + VV)$ & 98\% / 98\% & 2.74/3.15\% \\

$22.5^\circ$/0.5 & $\frac{1}{\sqrt{2}}(HH - iVV)$ & 98\% / 99\%
& 27.7/28.1\% & $-67.5^\circ$ & $\frac{1}{\sqrt{2}}(HH + VV)$ & 97\% / 99\% & 2.95/2.4\%\\

$45^\circ$/1.0  & $\frac{1}{\sqrt{2}}(HH + VV)$   & 98\% / 99\%
& 5.6/2.64\% & $-45^\circ$   & $\frac{1}{\sqrt{2}}(HH + VV)$ & 98\% / 99\% & 3.3/2.63\%\\

$67.5^\circ$/1.5 & $\frac{1}{\sqrt{2}}(HH + iVV)$ & 97\% / 98\%
& 28.2/28.7\% & $-22.5^\circ$ & $\frac{1}{\sqrt{2}}(HH + VV)$ & 96\% / 98\% & 2.91/2.93\%\\

$90^\circ$/2.0  & $\frac{1}{\sqrt{2}}(HH - VV)$   & 98\% / 98\%
& 2.76/3.183\% & $0^\circ$     & $\frac{1}{\sqrt{2}}(HH + VV)$ & 98\% / 98\% & 2.75/2.96\%\\

\bottomrule
\end{tabular}
\label{Table1}
\end{table*}

\section{Conclusions}

In conclusion, we have investigated both theoretically and experimentally the effect of the relative phase of Bell states in an entanglement-based QKD system. By introducing a geometric phase at the pump, we demonstrated deterministic transformation of a Bell state into a phase-shifted Bell state, leading to a periodic variation of the QBER. Although all phase-shifted Bell states are maximally entangled, their suitability for QKD depends critically on the relative phase. We find that the QBER remains below the security threshold only for relative phases within $\pm 0.31\pi$, beyond which secure key distribution is not possible. Even within this range, the tolerance to external perturbations is limited. We further show that the relative phase, whether introduced deliberately or arising from experimental imperfections such as crystal displacement, can be accurately measured and efficiently compensated using a geometric phase at the detection stage. This compensation restores the Bell state and maintains a consistently low QBER ($\sim3\%$), independent of the initial phase. Quantum state tomography confirms high-fidelity recovery of the Bell states, with fidelity exceeding $98\%$. These results establish geometric phase as a practical single-parameter tool for both engineering and correcting entangled states, significantly relaxing the stringent requirements on source optimization and enabling robust, high-quality entanglement for realistic QKD implementations. Finally, the geometric phase control and phase-noise compensation scheme demonstrated here can be extended to time-bin QKD by mapping temporal modes to polarization. This single-parameter phase tuning, combined with active feedback, can serve as a complementary approach to conventional interferometric stabilization and offers a promising route toward improved phase stability and low-QBER operation in practical quantum communication systems.

\section*{ACKNOWLEDGMENTS}
A. K. N. and G. K. S. acknowledge the support of the Department of Space, Govt. of India. A. K. N. acknowledges funding support for Chanakya - PhD fellowship from the National Mission on Interdisciplinary Cyber-Physical Systems of the Department of Science and Technology, Govt. of India through the I-HUB Quantum Technology Foundation.  G. K. S. acknowledges the support of the Department of Science and Technology, Govt. of India, through the Technology Development Program (Project DST/TDT/TDP-03/2022).

\section*{AUTHOR DECLARATIONS}
\subsection*{Conflict of Interest}
The authors have no conflicts to disclose.
\subsection*{Author Contributions}
A. K. N. developed the experimental setup and performed measurements. A. K. N. participated in experiments, data analysis, numerical simulation, and data interpretation. G. K. S. developed the ideas and led the project. All authors participated in the discussion and contributed to the manuscript writing.

\section*{DATA AVAILABILITY}
The data that support the findings of this study are available from the corresponding author upon reasonable request.

\bibliographystyle{IEEEtran}
\bibliography{Bib}

\begin{thebibliography}{10}
\providecommand{\url}[1]{#1}
\csname url@samestyle\endcsname
\providecommand{\newblock}{\relax}
\providecommand{\bibinfo}[2]{#2}
\providecommand{\BIBentrySTDinterwordspacing}{\spaceskip=0pt\relax}
\providecommand{\BIBentryALTinterwordstretchfactor}{4}
\providecommand{\BIBentryALTinterwordspacing}{\spaceskip=\fontdimen2\font plus
\BIBentryALTinterwordstretchfactor\fontdimen3\font minus \fontdimen4\font\relax}
\providecommand{\BIBforeignlanguage}[2]{{%
\expandafter\ifx\csname l@#1\endcsname\relax
\typeout{** WARNING: IEEEtran.bst: No hyphenation pattern has been}%
\typeout{** loaded for the language `#1'. Using the pattern for}%
\typeout{** the default language instead.}%
\else
\language=\csname l@#1\endcsname
\fi
#2}}
\providecommand{\BIBdecl}{\relax}
\BIBdecl

\bibitem{ekert1991}
A.~K. Ekert, ``Quantum cryptography based on bell’s theorem,'' \emph{Physical review letters}, vol.~67, no.~6, p. 661, 1991.

\bibitem{bennett1992}
C.~H. Bennett, G.~Brassard, and N.~D. Mermin, ``Quantum cryptography without bell’s theorem,'' \emph{Physical review letters}, vol.~68, no.~5, p. 557, 1992.

\bibitem{gisin2010}
N.~Gisin and R.~T. Thew, ``Quantum communication technology,'' \emph{Electronics letters}, vol.~46, no.~14, pp. 965--967, 2010.

\bibitem{Kwiat:99}
P.~G. Kwiat, E.~Waks, A.~G. White, I.~Appelbaum, and P.~H. Eberhard, ``Ultrabright source of polarization-entangled photons,'' \emph{Phys. Rev. A}, vol.~60, pp. R773--R776, Aug 1999.

\bibitem{kumar2025}
V.~Kumar, C.~Kaushik, M.~Ebrahim-Zadeh, C.~Chandrashekar, and G.~Samanta, ``Classical-to-quantum transfer of geometric phase for non-interferometric phase measurement and manipulation of quantum state,'' \emph{arXiv preprint arXiv:2505.20108}, 2025.

\bibitem{kim2006phase}
T.~Kim, M.~Fiorentino, and F.~N. Wong, ``Phase-stable source of polarization-entangled photons using a polarization sagnac interferometer,'' \emph{Physical Review A—Atomic, Molecular, and Optical Physics}, vol.~73, no.~1, p. 012316, 2006.

\bibitem{kim2019pulsed}
H.~Kim, O.~Kwon, and H.~S. Moon, ``Pulsed sagnac source of polarization-entangled photon pairs in telecommunication band,'' \emph{Scientific reports}, vol.~9, no.~1, p. 5031, 2019.

\bibitem{wang2011degree}
Y.~Wang, Y.~Zhang, J.~Wang, and J.~Jia, ``Degree of polarization for single-photon beam in a turbulent atmosphere,'' \emph{Optics Communications}, vol. 284, no.~13, pp. 3221--3226, 2011.

\bibitem{feng2024pre}
T.~Feng, R.~Fan, R.~Fan, C.~Wang, C.~Huang, A.~Li, H.~Zhang, Y.~Cao, and L.~Li, ``Pre-and post-compensation to suppress birefringent walk-off effects of entangled photons,'' \emph{Optics Express}, vol.~32, no.~23, pp. 40\,283--40\,292, 2024.

\bibitem{zhou2013ultra}
Z.-Y. Zhou, Y.-K. Jiang, D.-S. Ding, and B.-S. Shi, ``An ultra-broadband continuously-tunable polarization-entangled photon-pair source covering the c+ l telecom bands based on a single type-ii ppktp crystal,'' \emph{Journal of Modern Optics}, vol.~60, no.~9, pp. 720--725, 2013.

\bibitem{cialdi2011nonlocal}
S.~Cialdi, D.~Brivio, E.~Tesio, and M.~G. Paris, ``Nonlocal compensation of pure phase objects with entangled photons,'' \emph{Physical Review A—Atomic, Molecular, and Optical Physics}, vol.~84, no.~4, p. 043817, 2011.

\bibitem{steinlechner2013}
F.~Steinlechner, S.~Ramelow, M.~Jofre, M.~Gilaberte, T.~Jennewein, J.~P. Torres, M.~W. Mitchell, and V.~Pruneri, ``Phase-stable source of polarization-entangled photons in a linear double-pass configuration,'' \emph{Optics express}, vol.~21, no.~10, pp. 11\,943--11\,951, 2013.

\bibitem{kwiat1995new}
P.~G. Kwiat, K.~Mattle, H.~Weinfurter, A.~Zeilinger, A.~V. Sergienko, and Y.~Shih, ``New high-intensity source of polarization-entangled photon pairs,'' \emph{Physical Review Letters}, vol.~75, no.~24, p. 4337, 1995.

\bibitem{altepeter2005phase}
J.~B. Altepeter, E.~R. Jeffrey, and P.~G. Kwiat, ``Phase-compensated ultra-bright source of entangled photons,'' \emph{Optics Express}, vol.~13, no.~22, pp. 8951--8959, 2005.

\bibitem{feng2023ent}
T.~Feng, S.~Zhang, T.~Wu, Z.~Song, and L.~Li, ``Entangled photon-pair source using a wedge-shaped nonlinear crystal,'' \emph{Optical Materials}, vol. 145, p. 114441, 2023.

\bibitem{mo2005faraday}
X.-F. Mo, B.~Zhu, Z.-F. Han, Y.-Z. Gui, and G.-C. Guo, ``Faraday--michelson system for quantum cryptography,'' \emph{Optics letters}, vol.~30, no.~19, pp. 2632--2634, 2005.

\bibitem{zhou2010phase}
Z.-Q. Zhou, C.-F. Li, G.~Chen, J.-S. Tang, Y.~Zou, M.~Gong, and G.-C. Guo, ``Phase compensation enhancement of photon pair entanglement generated from biexciton decay in quantum dots,'' \emph{Physical Review A—Atomic, Molecular, and Optical Physics}, vol.~81, no.~6, p. 064302, 2010.

\bibitem{patel2016quantum}
R.~B. Patel, J.~Ho, F.~Ferreyrol, T.~C. Ralph, and G.~J. Pryde, ``A quantum fredkin gate,'' \emph{Science advances}, vol.~2, no.~3, p. e1501531, 2016.

\bibitem{tagliavacche2025}
N.~Tagliavacche, M.~Borghi, G.~Guarda, D.~Ribezzo, M.~Liscidini, D.~Bacco, M.~Galli, and D.~Bajoni, ``Frequency-bin entanglement-based quantum key distribution,'' \emph{npj Quantum Information}, vol.~11, no.~1, p.~60, 2025.

\bibitem{chen2008active}
W.~Chen, Z.~Han, X.~Mo, F.~Xu, G.~Wei, and G.~Guo, ``Active phase compensation of quantum key distribution system,'' \emph{Chinese Science Bulletin}, vol.~53, no.~9, pp. 1310--1314, 2008.

\bibitem{Lai:91}
G.~Lai and T.~Yatagai, ``Generalized phase-shifting interferometry,'' \emph{J. Opt. Soc. Am. A}, vol.~8, no.~5, pp. 822--827, May 1991.

\bibitem{Joenathan:94}
C.~Joenathan, ``Phase-measuring interferometry: new methods and error analysis,'' \emph{Appl. Opt.}, vol.~33, no.~19, pp. 4147--4155, Jul 1994.

\bibitem{wyant2003}
J.~C. Wyant, ``Dynamic interferometry,'' \emph{Optics and photonics news}, vol.~14, no.~4, pp. 36--41, 2003.

\bibitem{pancharatnam1956}
S.~Pancharatnam, ``Generalized theory of interference, and its applications: Part i. coherent pencils,'' in \emph{Proceedings of the Indian Academy of Sciences-Section A}, vol.~44, no.~5.\hskip 1em plus 0.5em minus 0.4em\relax Springer, 1956, pp. 247--262.

\bibitem{berry1987}
M.~V. Berry, ``The adiabatic phase and pancharatnam's phase for polarized light,'' \emph{Journal of Modern Optics}, vol.~34, no.~11, pp. 1401--1407, 1987.

\bibitem{simon1988}
R.~Simon, H.~Kimble, and E.~Sudarshan, ``Evolving geometric phase and its dynamical manifestation as a frequency shift: an optical experiment,'' \emph{Physical review letters}, vol.~61, no.~1, p.~19, 1988.

\bibitem{sjoqvist2008}
E.~Sj{\"o}qvist, ``A new phase in quantum computation,'' \emph{Physics}, vol.~1, p.~35, 2008.

\bibitem{zhou2025}
Z.~Zhou, S.~C. Carrasco, C.~Sanner, V.~S. Malinovsky, and R.~Folman, ``Geometric phase amplification in a clock interferometer for enhanced metrology,'' \emph{Science advances}, vol.~11, no.~18, p. eadr6893, 2025.

\bibitem{conrads2025}
L.~Conrads, F.~Bontke, A.~Mathwieser, P.~Buske, M.~Wuttig, R.~Schmitt, C.~Holly, and T.~Taubner, ``Infrared beam-shaping on demand via tailored geometric phase metasurfaces employing the plasmonic phase-change material in3sbte2,'' \emph{Nature Communications}, vol.~16, no.~1, p. 3698, 2025.

\bibitem{simon1989}
R.~Simon and N.~Mukunda, ``Universal su (2) gadget for polarization optics,'' \emph{Physics Letters A}, vol. 138, no.~9, pp. 474--480, 1989.

\bibitem{gisin2002quantum}
N.~Gisin, G.~Ribordy, W.~Tittel, and H.~Zbinden, ``Quantum cryptography,'' \emph{Reviews of modern physics}, vol.~74, no.~1, p. 145, 2002.

\bibitem{scarani2009}
V.~Scarani, H.~Bechmann-Pasquinucci, N.~J. Cerf, M.~Du{\v{s}}ek, N.~L{\"u}tkenhaus, and M.~Peev, ``The security of practical quantum key distribution,'' \emph{Reviews of modern physics}, vol.~81, no.~3, pp. 1301--1350, 2009.

\bibitem{Davis:25}
J.~J.~J. Davis, C.~L. Jackman, R.~Leonhardt, P.~J. Werbos, and M.~D. Hoogerland, ``Phase-shifted bell states,'' \emph{J. Opt. Soc. Am. B}, vol.~42, no.~6, pp. 1227--1235, Jun 2025.

\bibitem{Jabir:2017}
M.~Jabir, ``Robust, high brightness, degenerate entangled photon source at room temperature,'' \emph{Scientific reports}, vol.~7, no.~1, p. 12613, 2017.

\bibitem{Radhika:09}
R.~Rangarajan, M.~Goggin, and P.~Kwiat, ``Optimizing type-i polarization-entangled photons,'' \emph{Opt. Express}, vol.~17, no.~21, pp. 18\,920--18\,933, Oct 2009.

\bibitem{clauser1969}
J.~F. Clauser, M.~A. Horne, A.~Shimony, and R.~A. Holt, ``Proposed experiment to test local hidden-variable theories,'' \emph{Physical review letters}, vol.~23, no.~15, p. 880, 1969.

\bibitem{weihs1998violation}
G.~Weihs, T.~Jennewein, C.~Simon, H.~Weinfurter, and A.~Zeilinger, ``Violation of bell's inequality under strict einstein locality conditions,'' \emph{Physical Review Letters}, vol.~81, no.~23, p. 5039, 1998.

\bibitem{scheidl2009feasibility}
T.~Scheidl, R.~Ursin, A.~Fedrizzi, S.~Ramelow, X.-S. Ma, T.~Herbst, R.~Prevedel, L.~Ratschbacher, J.~Kofler, T.~Jennewein \emph{et~al.}, ``Feasibility of 300 km quantum key distribution with entangled states,'' \emph{New Journal of Physics}, vol.~11, no.~8, p. 085002, 2009.

\bibitem{neumann2021model}
S.~P. Neumann, T.~Scheidl, M.~Selimovic, M.~Pivoluska, B.~Liu, M.~Bohmann, and R.~Ursin, ``Model for optimizing quantum key distribution with continuous-wave pumped entangled-photon sources,'' \emph{Physical review A}, vol. 104, no.~2, p. 022406, 2021.

\bibitem{Singh:2023}
S.~Singh, V.~Kumar, A.~Ghosh, A.~Forbes, and G.~K. Samanta, ``A tolerance-enhanced spontaneous parametric downconversion source of bright entangled photons,'' \emph{Advanced Quantum Technologies}, vol.~6, no.~2, p. 2200121, 2023.

\end{thebibliography}

\end{multicols}
\end{document}